\documentclass[%
 aip,
 amsmath,amssymb,
 preprint,%
]{revtex4-1}

\newcommand*{\hh}{H$_2$~}
\newcommand*{\etal}{\textit{et al.}~}

\usepackage{graphicx}% Include figure files
\usepackage{dcolumn}% Align table columns on decimal point 
\usepackage{bm}% bold math
\usepackage[utf8]{inputenc}
\usepackage[T1]{fontenc}
\usepackage{mathptmx}
\usepackage{etoolbox}
\usepackage{color}

\makeatletter
\def\@email#1#2{%
 \endgroup
 \patchcmd{\titleblock@produce}
  {\frontmatter@RRAPformat}
  {\frontmatter@RRAPformat{\produce@RRAP{*#1\href{mailto:#2}{#2}}}\frontmatter@RRAPformat}
  {}{}
}%
\makeatother

\usepackage{xr}

\begin{document}

\title[]{Mechanisms of adsorbing hydrogen gas on metal decorated graphene}

\author{Yasmine S. Al-Hamdani}
\affiliation{Department of Earth Sciences, University College London, London WC1E 6BT, United Kingdom}
\affiliation{Thomas Young Centre, University College London, London WC1E 6BT, United Kingdom}
\affiliation{London Centre for Nanotechnology, University College London, London WC1E 6BT, United Kingdom}
\author{Andrea Zen}%
\affiliation{Dipartimento di Fisica Ettore Pancini, Università di Napoli Federico II, Monte S. Angelo, I-80126 Napoli, Italy}
\author{Angelos Michaelides}%
\affiliation{Yusuf Hamied Department of Chemistry, University of Cambridge, Cambridge CB2 1EW, United Kingdom}
\author{Dario Alf\`e}
\affiliation{Dipartimento di Fisica Ettore Pancini, Università di Napoli Federico II, Monte S. Angelo, I-80126 Napoli, Italy}
\affiliation{Department of Earth Sciences, University College London, London WC1E 6BT, United Kingdom}
\affiliation{Thomas Young Centre, University College London, London WC1E 6BT, United Kingdom}
\affiliation{London Centre for Nanotechnology, University College London, London WC1E 6BT, United Kingdom}

\date{\today}

\begin{abstract}
Hydrogen is a key player in global strategies to reduce greenhouse gas emissions. In order to make hydrogen a widely-used fuel, we require more efficient methods of storing it than the current standard of pressurized cylinders. An alternative method is to adsorb \hh in a material and avoid the use of high pressures. Among many potential materials, layered materials such as graphene present a practical advantage as they are lightweight. However, graphene and other 2D materials typically bind \hh too weakly to store it at the typical operating conditions of a hydrogen fuel cell. Modifying the material, for example by decorating graphene with adatoms, can strengthen the adsorption energy of \hh molecules, but the underlying mechanisms are still not well understood. 
In this work, we systematically screen alkali and alkaline earth metal decorated graphene sheets for the adsorption of hydrogen gas from first principles, and focus on the mechanisms of binding. We show that there are three mechanisms of adsorption on metal decorated graphene and each leads to distinctly different hydrogen adsorption structures. The three mechanisms can be described as weak van der Waals physisorption, metal adatom facilitated polarization, and Kubas adsorption. Among these mechanisms, we find that Kubas adsorption is easily perturbed by an external electric field, providing a way to tune \hh adsorption.
\end{abstract}

\maketitle

\section{Introduction}

%motivation
There is an urgent need to reduce the use of fossil fuels and develop alternative, less polluting, methods of energy production. To this end, \hh is long-standing potential candidate fuel.\cite{Allendorf2018} There is an energy cost to producing \hh in the first place, but \hh molecules provide almost three times the energy density by weight as fossil fuels\cite{Armaroli2011} and burning \hh produces water with no additional harmful pollutants. In addition to burning, hydrogen can be combined with oxygen more efficiently in fuel cells, producing electricity and still only water as waste. At present, H$_2$ is stored as pressurised gas and more efficient \hh storage materials are needed to propel this fuel into wide-scale use.

%background
A promising method of storing hydrogen fuel is to physisorb \hh molecules in a lightweight material. Cycling weakly adsorbed hydrogen gas through a material is expected to have minimal degradation effect on the storage material as \hh molecules remain intact. Other adsorption mechanisms of storage, such as the spillover method, rely on \hh dissociating and forming covalent bonds with the storage material which makes the material more susceptible to deformation. 
In addition, weakly adsorbed hydrogen molecules require less energy to be released from a material relative to chemisorbed hydrogen atoms. 
% Explain 200-400 meV window
The window for ideal \hh adsorption energy can be estimated in a heuristic approach and considering the typical working temperature and pressure of fuel cells. The pressure ($\mathrm{p}$), temperature ($\mathrm{T}$), and the adsorption energy ($\mathrm{E_{ads}}$), can be approximately related according to: 
\begin{equation}\label{eq:vp-t}
p = { e^{\frac {E_{ads}}{k_BT} } }  
    {\left(2 \pi {m_{H_2}}\right)^{3\over2}  \over h^3} 
    \left({k_BT}\right)^{5\over2} 
    2 \sinh\left( {\hbar\omega_z\over{2k{_B}T}} \right)
\end{equation}
where ($\mathrm{k_B}$) is the Boltzmann constant, $h$ is Planck's constant,  $m_{H_2}$ is the mass of H$_2$, $I$ is the moment of inertia, $\omega_z$ is the harmonic frequency of vibration of the H$_2$ molecule with respect to the substrate. For a full account of how Eq.~\ref{eq:vp-t} is used and the approximations we make, see the Appendix.

\begin{figure}[ht]
    \centering
    \includegraphics[width=0.5\textwidth]{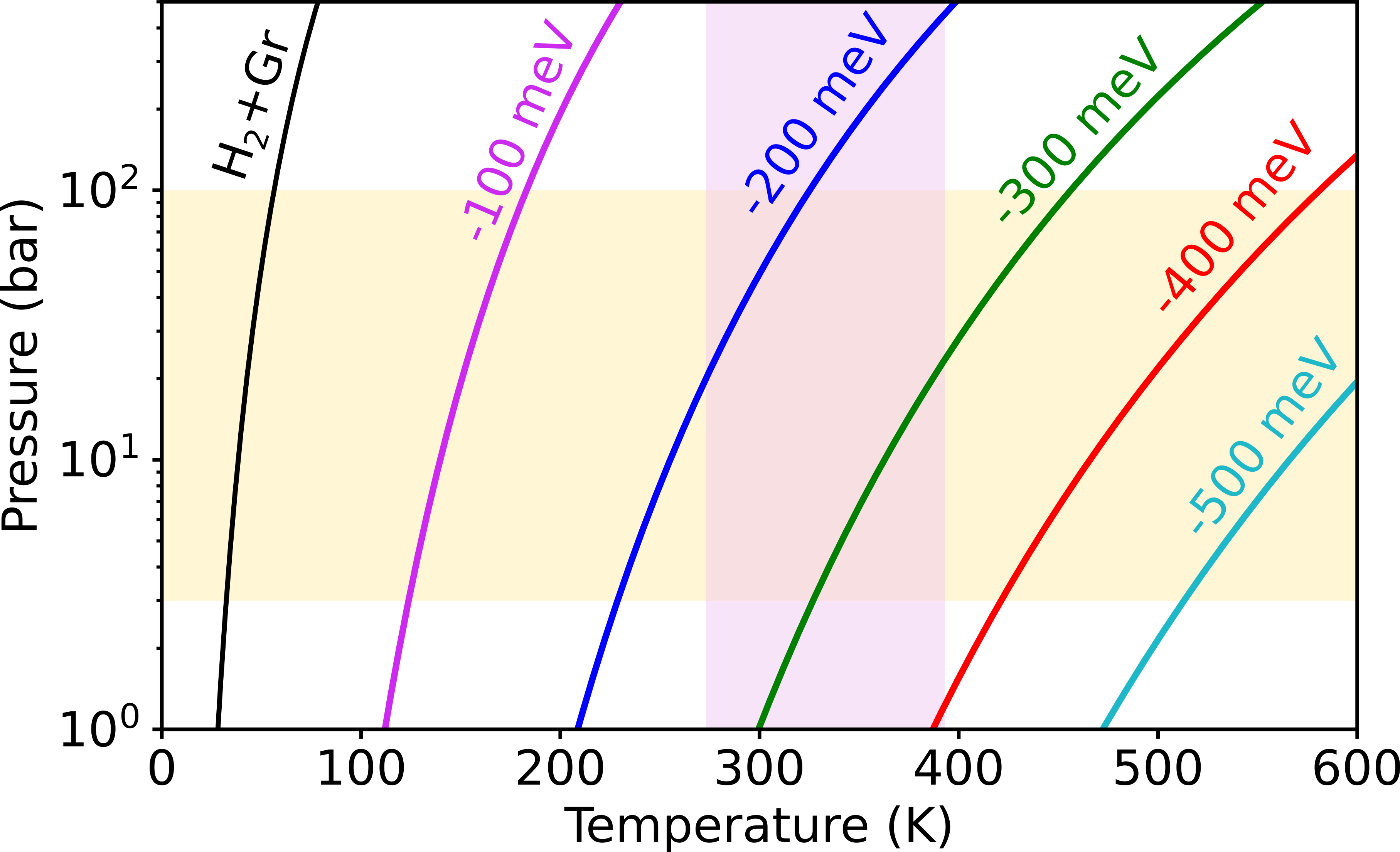}
    \caption{The temperature-pressure profile of \hh on pristine graphene at different adsorption energies indicated by the colored lines. The black line corresponds to the reference \hh adsorption energy ($-24\pm11$ meV) on pristine graphene from diffusion Monte Carlo.\cite{al-hamdani2017cnt} An ideal range of \hh vapor pressures for a typical fuel cell is indicated by the horizontal yellow region and the ideal working temperatures are indicated by the vertical pink region. The overlap in ideal temperature and pressure is roughly bounded by \hh adsorption energies of $-200$ and $-400$ meV. See Eq.~\ref{eq:vp-t} for the relation between pressure, temperature and adsorption energy.}  
    \label{fig:pT}
\end{figure}

Polymer electrolyte membrane (PEM) fuel cells have been developed for a range of operating temperatures, with high temperature PEM fuel cells functioning above  370 K.\cite{Ogungbemi2021,Salam2020} Taking into account intermediate and high temperature PEM fuel cells, we consider working temperatures of 270-390 K in this work. The typical operating pressure of a PEM fuel cell is $\sim3$ bars of \hh pressure\cite{Ogungbemi2021} which means that the storage material must have a higher \hh vapor pressure to readily release \hh to the fuel cell. In addition, an upper-bound of 100 bar has been proposed for the \hh vapor pressure to avoid similar technological challenges as containing a highly pressurized gas~\cite{Ahmed2019}.
Under such conditions, it is generally accepted that the free energy of \hh adsorption in a material is $-200$ to $-400$ meV per \hh molecule as can be seen from Fig.~\ref{fig:pT}. The challenge of finding a material that binds \hh suitably is also exacerbated by additional factors such as the weight and volume of the storage material.\cite{Armaroli2011,Allendorf2018,Lai2019} Evidently, lighter and low-volume materials are required for practical and energy efficient fuel storage for mobile applications.  

There are various promising materials for \hh storage and among them, we are interested in layered materials, such as graphene, as they are lightweight and are able to adsorb molecular hydrogen. However, the adsorption energy of \hh on pristine graphene is predicted to be less than $-50$ meV\cite{al-hamdani2017cnt} which is too weak for viable hydrogen storage (see Fig.~\ref{fig:pT}). Structural defects and decoration by adatoms is known to enhance the adsorption energy of molecules on graphene\cite{Spyrou2013} and there are countless combinations that can be considered. However, it is experimentally challenging to produce well controlled and characterized graphene with defects or adatoms and therefore it is difficult to ascertain the \hh storage capacity of such potentially useful materials.    

%the challenge
To date, there have been indications that decorating graphene with alkali and alkaline earth metal adatoms facilitates \hh adsorption,\cite{Wong2014,Reunchan2011,Wen2017,Amaniseyed2019,Sun2010,Beheshti2011,Ataca2009,Zhou2012} potentially yielding adequate \hh capacities by weight. However, experimental information is scarce and computational efforts to understand \hh adsorption on metal decorated graphene are difficult to unify. For example, in different studies \hh adsorption energies have been predicted using different density functional approximations preventing us from drawing reliable trends. In addition, the structure of \hh molecules adsorbed around different metal adatoms on graphene has not received systematic focus and stands to be better understood. 
%

%Ca@Gr
Among alkali and alkaline earth metals, Ca decorated graphene is one of the most studied systems.\cite{Ataca2009,Beheshti2011,Wong2014,Wen2017,Reunchan2011,Cha2011} This is partly due to favorable \hh adsorption energies being predicted on this material as well as the relatively low cohesive energy of Ca, which is expected to prevent agglomeration on graphene.
Specifically, Ataca \textit{et al.} suggested over a decade ago that Ca adatoms facilitate the adsorption of \hh molecules via Kubas-type binding.\cite{Ataca2009} This mechanism involves stabilizing the $3d$ state of Ca relative to $4s$ and donating electron density from $3d$ into the H$_2$ $1\sigma^{*}$ state.\cite{Kubas2001} Since then, a number of wavefunction based methods have been used to understand the Ca$^+$-4H$_2$ cluster (without a graphene substrate) and deduce whether a Ca adatom is able to bind \hh using the Kubas mechanism.\cite{Cha2009,Bajdich2010,Sun2010,Cha2011,Park2012,Purwanto2011,Ohk2010} The general conclusion from these works is that Ca is unlikely to bind \hh using a Kubas-type binding and hence, cast doubt on the accuracy of density functional theory (DFT) approximations. However, graphene has been shown to affect adsorption and importantly, some metal adatoms (including Ca) make the adatom-graphene system metallic. Therefore, it is not straightforward to infer the nature of interaction on graphene from predictions on gas phase clusters.

Alongside Ca, other alkali and alkaline earth metals on graphene have been considered for \hh adsorption.\cite{Zhou2012,Amaniseyed2019,Reunchan2011,Wen2017,Wong2014} In brief, previous works have focused on assessing the adsorption strength of \hh on a given material %without addressing the adsorption configurations carefully and 
and in some cases methods without dispersion were used to predict adsorption energies.\cite{Reunchan2011,Zhou2012} We seek to build a better understanding of the mechanisms underpinning \hh adsorption on different alkali and alkaline earth metal adatoms on graphene. 
In this work, we systematically compute \hh adsorption on alkali and alkaline earth metal decorated graphene and draw mechanistic insights. We outline our computational setup and methods in Section~\ref{sec:methods}. In Section~\ref{sec:screening} we report the results of screening 1 to 7 \hh molecules per metal adatom on graphene. We refine and analyze the adsorption of \hh for a subset of systems in Section~\ref{sec:mechanisms}. In doing so, we elucidate the mechanisms of adsorption and find that they can be summarized in three physically distinct categories. In Section~\ref{sec:efield}, we report the effect of applying an external electric field on the \hh interaction with the substrate and find that it depends strongly on the binding mechanism. We conclude in Section~\ref{sec:conclusions} with a brief discussion of the results.

\section{Methods}\label{sec:methods}
The initial screening of adsorption energies was performed with CP2K v.7.1\cite{Kuhne2020,Vandevondele2005} and Goedecker-Teter-Hutter pseudopotentials\cite{Goedecker1996,Krack2005} in combination with DZVP-MOLOPT-SR-GTH basis sets.\cite{VandeVondele2007} A maximum plane-wave cut-off of 300 Ry was used across 5 grids, with a relative cut-off of 30 Ry. Our CP2K calculations were spin-polarised and performed at $\Gamma$-point only for a ($5\times5$) unit cell of graphene. The geometries were optimized with the BFGS method until the maximum force was less than 5$\times10^{-4}$ Ha $a_{0}^{-1}$. All parameters of the CP2K geometry optimizations can be seen in the example input in the SM. The Perdew-Burke-Ernzerhof (PBE) exchange-correlation functional\cite{Perdew1996} was used in combination with Grimme's D3 dispersion method\cite{Grimme2010} with zero-type damping and three-body Axilrod-Teller-Muto terms included, to account for van der Waals interactions. 
% XC functional 
It is known that the choice of exchange-correlation functional has a notable impact on the H$_2$ adsorption energy on graphene-type surfaces.\cite{al-hamdani2017cnt,Wong2014} Particularly in the case of physisorption, long-range dispersion interactions are expected to play an important role and therefore it is necessary to use a dispersion method. However, in the absence of experimental reference adsorption energies for the systems we are considering, it is difficult to ascertain which dispersion method yields the most accurate results. In general, dispersion methods have been shown to predict consistent structures and relative energies.\cite{Bucko2010,Carter2014,Bedolla2014,Klimes2012,Carrasco2013,Rosa2014} Absolute adsorption energies, on the other hand, can vary considerably among different density approximations. Previously, we established diffusion Monte Carlo (DMC) reference adsorption energies for \hh inside and outside a carbon nanotube (CNT) and found that add-on dispersion methods are more accurate than seamless density-dependent dispersion functionals for the adsorption of H$_2$ inside a carbon nanotube.\cite{al-hamdani2017cnt} 
Add-on dispersion methods include the D3,\cite{Grimme2010} D4,\cite{Caldeweyher2017} and many-body-dispersion (MBD)\cite{Ambrosetti2014,Tkatchenko2012} methods. These partially account for beyond two-body dispersion interactions which can play an important role in graphene-like materials.\cite{Gobre2013} In our work, we combine results from two DFT packages and therefore, to be consistent, we use PBE+D3 as it is implemented in CP2K and VASP. Note that PBE+MBD and PBE+D3 both predict an \hh adsorption energy of $-53$ meV on pristine graphene, while DMC yields $-24 \pm 11$ meV.\cite{al-hamdani2017cnt}  

Metal decorated graphene (M@Gr) was modelled using a ($5\times5$) unit cell of graphene with unit cell parameters optimized using PBE+D3. A single metal atom (M) was placed at the hollow site and fully optimized for Li, Be, Na, Mg, K, Ca, Rb, Sr, Cs, and Ba. H$_2$ molecules were placed upright relative to graphene and surrounding the metal atom in every initial structure. An inter-layer spacing of 20~\AA\ is applied along the $z$-axis between graphene sheets and dipole corrections\cite{Neugebauer1992,Makov1995} along $z$-direction also computed. Up to 7 H$_2$ molecules were fully optimized on each M@Gr system, totalling 70 systems, with all atoms in the cell allowed to relax. We report the results of this screening in Section \ref{sec:screening}.

For a better understanding of the binding mechanisms and to assess the quality of the initial screening, we performed fine-grained optimizations of the resulting geometries from the screening. We used VASP v.5.4.4\cite{Kresse1993,Kresse1994,Kresse1996,Kresse1996a} with standard PAW potentials and a 500 eV plane-wave cut-off. Since the neutral metal atoms are easily ionized, potentials with explicit semi-core $s$ states were used for all metals. Na has the highest energy core states amongst the metal atoms we considered and we found that the interaction energy of 4H$_2$ on Na@Gr is converged with a 500 eV plane-wave cut-off to within 2 meV. 
In addition, the decoration of graphene with metal atoms makes the system metallic and hence we used a dense \textbf{k}-point mesh of $9\times9\times1$ centred on $\Gamma$. We found the interaction energy of 4H$_2$ on Ca@Gr is converged within 1 meV per H$_2$ using a \textbf{k}-point mesh of $5\times5\times1$ and therefore we expect an even denser mesh to be sufficient for all the systems we considered. The fine-grained geometry relaxations for 3-5 \hh molecules on each substrate were converged with residual forces less than 0.01 eV \AA$^{-1}$. Densities of states were obtained using a $15\times15\times1$ \textbf{k}-point mesh and the SUMO code\cite{Ganose2018} was used in post-processing the data. For the application of external electric force fields in Section~\ref{sec:efield}, we used a sawtooth potential as implemented in VASP and applied the field along the z-direction in the unit cell, \textit{i.e.} perpendicular to the graphene sheet. We also performed geometry optimizations of 4\hh adsorbed on Ca@Gr at two electric fields (0.2 V \AA$^{-1}$ and $-0.2$ V \AA$^{-1}$) using a \textbf{k}-point mesh of $5\times5\times1$.

%\section{Results}
\section{Screening H$_2$ adsorption on metal decorated graphene}\label{sec:screening}

\begin{figure}[ht]
    \centering
    \includegraphics[width=1\textwidth]{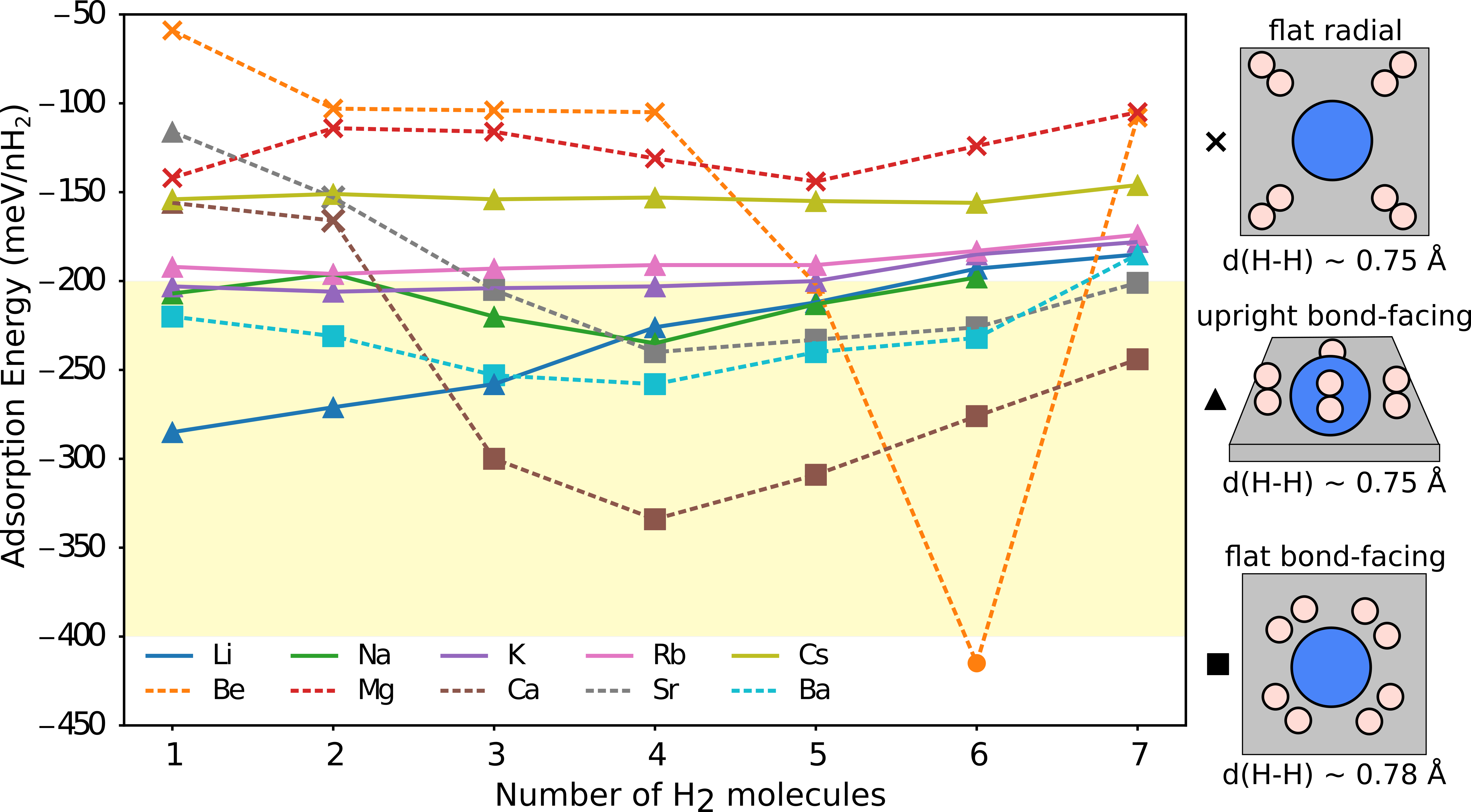}
    \caption{Preliminary screening of H$_2$ adsorption on group 1 (solid lines) and group 2 (dashed lines) M@Gr. PBE+D3 adsorption energies shown here (in meV) are approximate only. For a converged PBE+D3 adsorption energies, see Table 1 where a subset of systems are reported. The symbols indicate the optimized orientation of \hh molecules around the metal atom. Triangles indicate upright bond-facing (BF), squares indicate flat BF, and crosses indicate flat radial configuration. The three mechanisms are also depicted on the right. The circle (6H$_2$+Be@Gr) indicates dissociative adsorption of \hh has occurred. Average H-H bond lengths are also given for each mechanism of binding. The yellow shaded region from $-200$ to $-400$ meV indicates an estimated range of suitable adsorption energies for storage in operation with fuel cells.}
    \label{fig:screen}
\end{figure}

Decorating graphene with single metal atoms has previously been found to strengthen the adsorption of \hh molecules for some metals such as Ca and Li.\cite{Ataca2009,Zhou2012} In some cases, such as Mg@Gr, the adsorption of \hh remains weak.\cite{Amaniseyed2019} We focus specifically on alkali and alkaline earth metals, from Li to Ba, with the aim to understand the mechanisms underpinning the interactions. The indication from previous works is that dispersion interactions contribute significantly to the adsorption energy\cite{Wong2017} and \hh is bound too weakly to be useful for hydrogen storage.\cite{Singh2012,Spyrou2013}
However, it appears from the range of adsorption energies reported, that it is difficult to establish consistent adsorption energies from DFT approximations.\cite{Singh2012} Moreover, a systematic analysis of the adsorption geometries is missing from our current understanding and we aim to address that here. An approximate overview of the relative strength of \hh adsorption as the number of \hh molecules are increased is given by the crude screening in this section. The results of the rapid DFT screening of H$_2$ adsorption energies on M@Gr is shown in Fig.~\ref{fig:screen}. The adsorption energy ($E_{ads}$) is defined as:
\begin{equation}\label{eq:ads}
    E_{ads} = (E^{tot}_{nH_2+M@Gr} - E^{tot}_{M@Gr} - nE^{tot}_{H_2} ) / n
\end{equation}
where $E^{tot}_{nH_2+M@Gr}$ is the total energy of \hh molecules adsorbed on M@Gr, $E^{tot}_{M@Gr}$ is the total energy of the fully relaxed M@Gr substrate, $E^{tot}_{H_2}$ is the total energy of the gas phase relaxed \hh molecule, and $n$ is the number of \hh molecules adsorbed. 

Screening calculations were performed at the $\Gamma$-point only and using atom-centered basis sets without correcting for basis set superposition error. As a result, the PBE+D3 adsorption energies in Fig.~\ref{fig:screen} are likely to be overestimated. For reliable PBE+D3 adsorption energies see Table 1  where we report adsorption details for systems with 3-5 H$_2$ molecules as well as the adsorption energy of metal adatoms on graphene. 

The geometry optimization of H$_2$ molecules on M@Gr broadly yields three orientations of H$_2$ molecules, as can be seen from Fig.~\ref{fig:screen}. There are several features to note from these preliminary adsorption profiles. 
First, the weakest adsorption profile is seen for Be@Gr and Mg@Gr, where the H$_2$ molecules prefer to be flat on the graphene sheet and pointing radially to the metal atom. An example of this flat radial configuration is illustrated in Fig.~\ref{fig:screen}. 
% previous study
This configuration suggests the main contribution to adsorption is between \hh and graphene, with an additional weak interaction with the metal adatom. Note that Be has a degeneracy in its valence states that is known to make it reactive with hydrogen, forming Be--H bonds. This occurs in one of the geometry optimizations, when 6 H$_2$ molecules are placed near Be, leading to the dissociative adsorption of a \hh molecule. Therefore, Mg and Be are not likely to be suitable adatoms on graphene for \hh storage via weak adsorption.  
Second, all alkali M@Gr substrates adsorb \hh in the upright bond-facing (BF) configuration and the adsorption energy profile is near-linear with increasing number of molecules. For K, Rb, and Cs, the adsorption profile is particularly flat, varying by less than 30 meV in the adsorption energy per \hh molecule, from 1 to 7 \hh molecules. Adsorption is strongest among alkali metals for Li@Gr with up to 3 \hh molecules. However the \hh adsorption energy on Li@Gr shows a steady weakening with increasing number of \hh molecules. This is due to \hh molecules not fitting around the small Li adatom and therefore spreading further away on the surface. In the case of Na@Gr, there is a small $\sim40$ meV variation in the \hh adsorption energy, with the most favorable binding occurring at 4\hh molecules. However, the configurations remain upright BF across the profile. 
We can see from Fig.~\ref{fig:screen} that another configuration of \hh (flat BF) results on Ca, Sr, and Ba, decorated graphene. The flat BF configuration is not exclusive on these substrates and both upright BF and flat radial configurations can be seen for 1, 2, and 7 \hh adsorbed molecules. Indeed, these heavier alkaline earth elements exhibit the most variation in their \hh adsorption profiles, varying by more than 70 meV with respect to the number of \hh molecules adsorbed. However, the strongest adsorption for graphene decorated with Ca, Sr, and Ba, is consistently predicted at 4 \hh molecules in the flat BF orientation. In addition, the flat BF configurations of \hh have a distinct H--H bond length of 0.78 \AA, \textit{i.e.} a $4\%$ elongation with respect to the equilibrium bond length. 
On the other hand, in the flat radial and upright BF configurations the H--H bond length stays close to equilibrium (0.75 \AA). The longer bond length for the flat BF configuration of \hh is therefore indicative of a different interaction mechanism that involves the $1\sigma^{*}$ state of the \hh molecule. This is known as the Kubas type bonding interaction and it has been discussed in previous works that considered Ca adatoms.\cite{Singh2012,Ataca2009,Cha2011,Sun2010} Here, we see that this configuration manifests more generally when graphene is decorated with alkaline earth metals that have available $d$-states, such as Sr and Ba. We also describe this mechanism in more detail in Section~\ref{sec:mechanisms}.

Our screening of \hh adsorption on alkali and alkaline earth metal decorated graphene suggests that the strongest non-dissociative adsorption of \hh for more than 3 molecules per adatom, occurs on Ca, Sr, and Ba decorated graphene. For less than 3 \hh molecules per adatom, Li@Gr is predicted to bind \hh strongly. However, adsorption energies in this screening are only approximate as loose technical parameters have been used and the PBE+D3 method is also a source of uncertainty. In the next section we report adsorption energies from well-converged geometry optimizations for a subset of systems with PBE+D3. 

\section{Mechanism of adsorption and the role of graphene}\label{sec:mechanisms}

To understand the electronic structure mechanisms underlying the three distinct configurations of \hh adsorption we find, we performed well-converged geometry relaxations on all adatom systems with 3-5 \hh molecules from Section~\ref{sec:screening}. The computational details are given in Section \ref{sec:methods} and we note that the main improvement is in the \textbf{k}-mesh density (using a $9\times9\times1$ grid on a $(5\times5)$ unit cell of graphene). We have also performed calculations with alternative starting geometries to see if flat BF configurations can be stabilized over upright BF on alkali metals, and \textit{vice versa} on alkaline earth metals. We find that the orientation of \hh molecules predicted in Section \ref{sec:screening} is consistent and that the graphene-adatom distances change by less than $5\%$ or 0.16 \AA~. Similarly, the H$_2$-adatom distance changes by, at most, 10$\%$ or $0.28$ \AA~ (more details provided in the SM).

The PBE+D3 \hh adsorption energies on M@Gr substrates and metal adatom adsorption energies on graphene are reported for well-converged optimized structures in Table 1.  The PBE+D3 metal adatom adsorption energies ($\mathrm{E_{M@Gr}}$) show that Mg and Be adsorb weaker than $-300$ meV on graphene, while other metal adatoms adsorb by over $-700$ meV. 
The average H$_2$-metal adatom and graphene-metal adatom separation distances are also reported in Table 1 for each metal considered. We can see that stronger \hh adsorption is accompanied by shorter H$_2$-metal adatom separation distances and that Ca and Ba adatoms best facilitate the adsorption of \hh molecules with adsorption energies of up to $-190$ and $-181$ meV per \hh molecule, respectively. It is evident that the screening in Section \ref{sec:screening} led to overestimated adsorption energies, but we note that the most favorable adsorption energy predicted here with PBE+D3 is within 10 meV of the range that is expected to be useful for \hh storage. 

Forming an understanding of the \hh adsorption mechanisms under idealized conditions may guide experiments towards realizing these systems and provides a basis for more accurate theoretical work. It is important to note, however, that the accuracy of PBE+D3 is not established for predicting M@Gr systems as there is no experimental or theoretical reference information. What is more, the zero-point energy contributions to the adsorption energy have not been taken into account and diffusion of molecules and atoms at the surface of graphene at finite temperatures is also likely to be an important factor in the ultimate use of such a material for \hh storage. To provide some indication of the stability of the materials predicted in this work, we computed 
%the barrier for the diffusion of Ca across graphene and 
the dissociation of \hh on Ca@Gr -- as it is the material with the strongest \hh adsorption among the systems considered. 
%First, we computed the barrier for Ca to diffuse from one hollow site to a next-nearest hollow site on pristine graphene using nudged elastic band (NEB) calculations\cite{Henkelman2000a,Henkelman2000b} (details are given in the SM). The PBE+D3 barrier is $\sim140$ meV, which is considerably higher than $\mathrm{k_B}$T at typical operating conditions for a PEM fuel cell and indicates that Ca will not diffuse on graphene. Second, 
\hh dissociating on Ca@Gr would indicate storage via the spillover effect instead and we gauge the likelihood of this by fully relaxing 2H+Ca@Gr, with H atoms chemisorbed on graphene in the vicinity of Ca for two configurations. The fully relaxed structures and computational details can be found in the SM. We find that 2H+Ca@Gr is $\sim1.7$ eV less stable than H$_2$+Ca@Gr, suggesting that intact H$_2$ is thermodynamically stable on Ca@Gr. These calculations provide preliminary indications, but further work is needed to cement our predictions.

\begin{table}
\caption{\label{tab:table1} Adsorption properties of 3-5 \hh molecules adsorbed on alkali and alkaline earth M@Gr from PBE+D3. $\mathrm{E_{M@Gr}}$ is the fully relaxed adsorption energy of the metal adatom (M) on a $(5\times5)$ unit cell of graphene (Gr) and $d_{\mathrm{M-Gr}}$ is the corresponding separation distance along the $z$-axis considering the average position of all carbon atoms. $\mathrm{E_{ads}^{nH_2}}$ is the average adsorption energy per \hh molecule when n\hh molecules are adsorbed (in eV). The H-H bond lengths, $d_{\mathrm{H-H}}$, and average M-\hh distances, $d_{\mathrm{M-H_{2}}}$, are reported for the 4H$_2$+M@Gr system in \AA. In the upper section, Li to Cs, an the \hh molecules are in an upright bond-facing. \hh molecules are in flat radial configuration on Be@Gr and Mg@Gr. In the lower section of the table, Ca to Ba, \hh molecules are in a flat bond-facing configuration. The values reported here correspond to spin-polarized geometry optimizations performed with $9\times9\times1$ \textbf{k}-point mesh and force convergence criterion of 0.01 eV \AA$^{-1}$. }
\begin{ruledtabular}
\begin{tabular}{lccccccc}
Adatom (M) & $\mathrm{E_{M@Gr}}$ (eV) & $d_{\mathrm{M-Gr}}$ (\AA) & $\mathrm{E_{ads}^{3H_2}}$ (eV) & $\mathrm{E_{ads}^{4H_2}}$ (eV) & $\mathrm{E_{ads}^{5H_2}}$ (eV) & $d_{\mathrm{H-H}}$ (\AA) & $d_\mathrm{M-H_{2}}$ (\AA) \\ \hline
Li         & $-1.279$  & 1.704         & $-0.187$          & $-0.161$           & $-0.141$    & 0.755     & 2.348   \\
Na         & $-0.719$  & 2.189         & $-0.176$          & $-0.173$           & $-0.156$    & 0.756     & 2.516   \\
K          & $-1.200$  & 2.571         & $-0.137$          & $-0.137$           & $-0.121$    & 0.754     & 2.964   \\
Rb         & $-1.262$  & 2.730         & $-0.128$          & $-0.128$           & $-0.112$    & 0.754     & 3.209   \\
Cs         & $-1.466$  & 2.903         & $-0.117$          & $-0.118$           & $-0.102$    & 0.753     & 3.435   \\ \hline
Be         & $-0.181$  & 3.218         & --                & $-0.088$           & --          & 0.754     & 2.875   \\
Mg         & $-0.281$  & 3.322         & --                & $-0.056$           & --          & 0.754     & 3.180   \\ \hline
Ca         & $-0.741$  & 2.314         & $-0.142$          & $-0.190$           & $-0.178$    & 0.784     & 2.287   \\
Sr         & $-0.753$  & 2.497         & $-0.096$          & $-0.135$           & $-0.132$    & 0.779     & 2.478   \\
Ba         & $-1.198$  & 2.577         & $-0.159$          & $-0.181$           & $-0.163$    & 0.771     & 2.722  
\end{tabular}
\end{ruledtabular}
\end{table}
\begin{figure}
    \centering
    \includegraphics[width=0.45\textwidth]{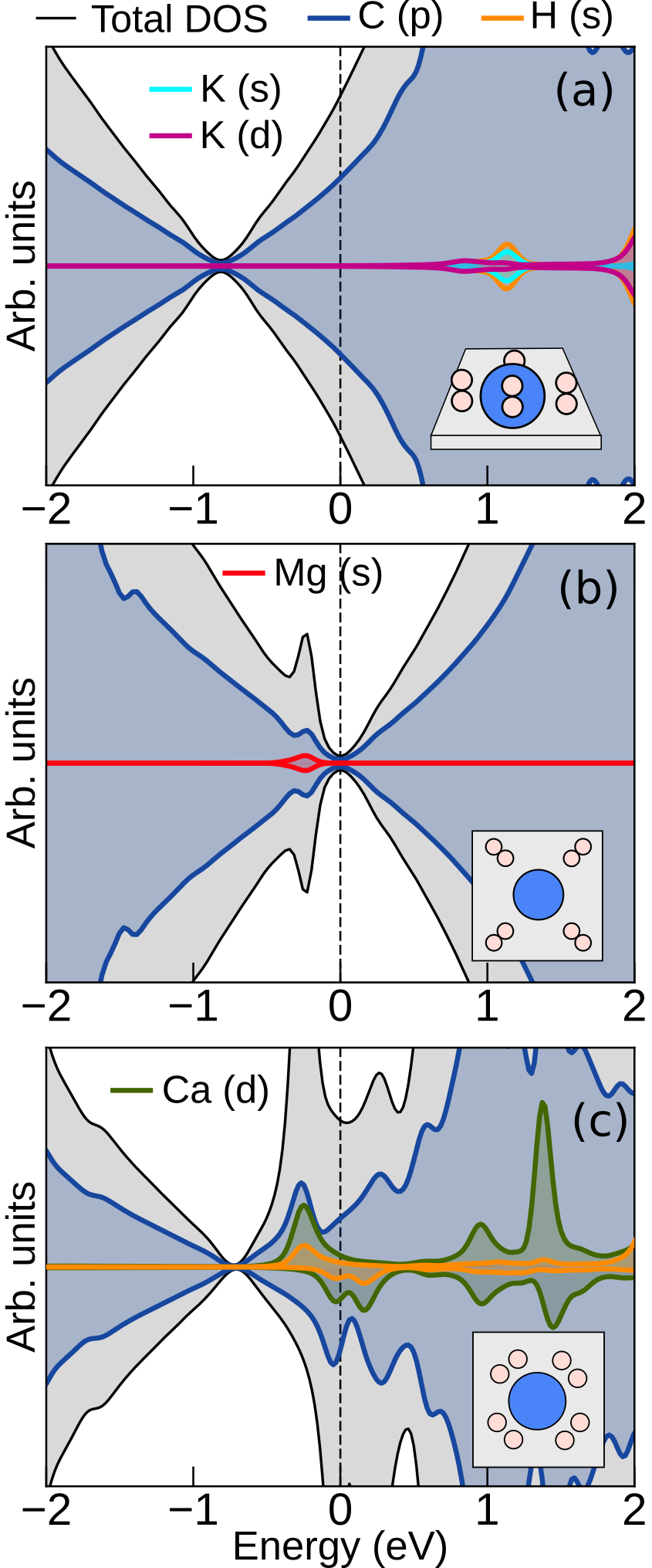}
    \caption{The projected density of states (PDOS) within $\pm$2 eV of the Fermi energy for 4H$_2$ adsorbed on (a) K@Gr, (b) Mg@Gr, and (c) Ca@Gr. The PDOS has been shifted to the Fermi energy for each system. The grey shaded region indicates the total DOS. H-$s$ projection shown in orange and C-$p$ projection shown in blue. H-$s$ near the Fermi energy is due to the $1\sigma^{*})$ state of H$_2$, while the $1\sigma$ state around $-8$ eV relative to the Fermi energy cannot be seen in this energy window. The projection is over spherical functions centred on the atoms and as such, the sum of projected states may not sum to the total DOS. A schematic of the configuration of 4\hh for each M@Gr system is shown in the insets.}
    \label{fig:mechdos}
\end{figure}

\begin{figure}
    \centering
    \includegraphics[width=1\textwidth]{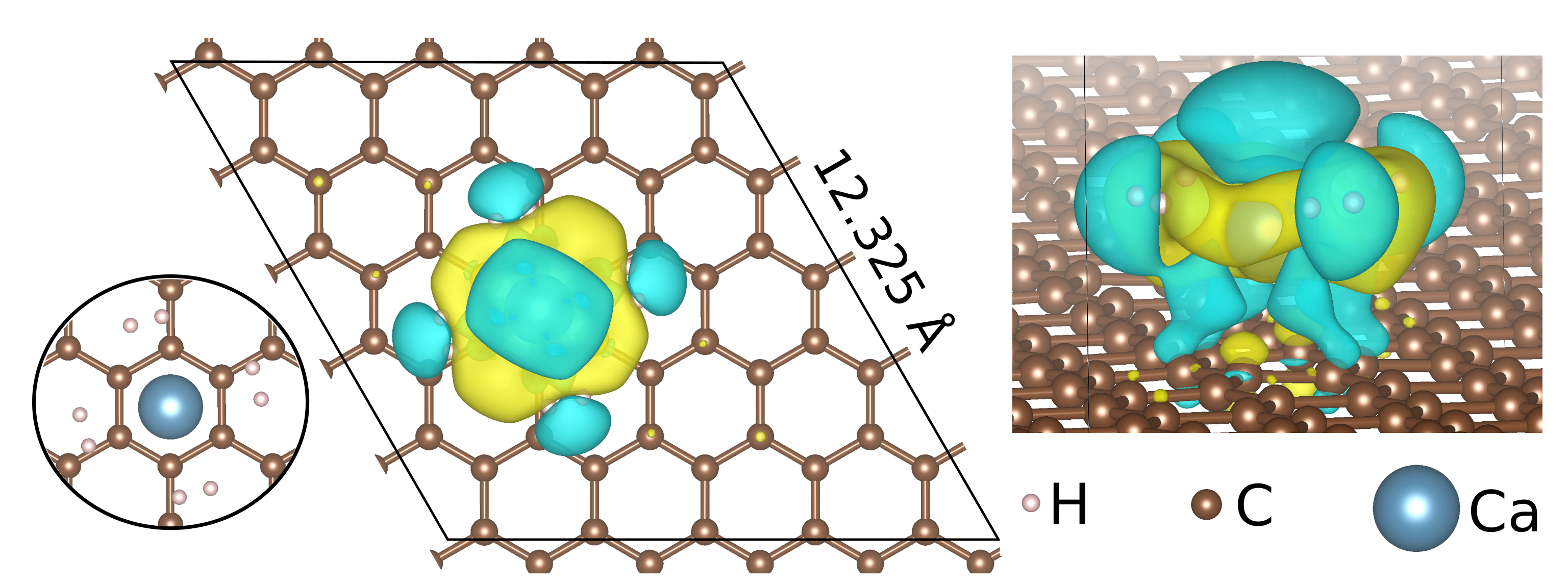}
    \caption{The 4H$_2$+Ca@Gr system showing the geometry of H$_2$ molecules around Ca and the charge redistribution upon adsorbing H$_2$ molecules. The unit cell used is indicated in the middle panel. Charge density difference is shown between 4H$_2$ and Ca@Gr using an isosurface level of 0.002 e \AA$^{-3}$. Charge density depletion is shown in blue and charge density accumulation is shown in yellow.}
    \label{fig:4h2cagr}
\end{figure}

First, we focus on the 4H$_2$+Ca@Gr system, where the adsorption energy is the strongest and there is a long-standing effort to establish whether the system is viable for \hh storage. The unit cell, adsorption configuration, and charge density difference for adsorbing 4\hh molecules can be seen in Fig.~\ref{fig:4h2cagr}. We can see that there is charge accumulation in the region between the Ca adatom and the \hh molecules and charge depletion above the Ca adatom and within the H--H bonding regions, in agreement with the work of Ataca \etal\cite{Ataca2009} Charge depletion along \hh covalent bonds is consistent with longer H--H bond lengths, from 0.75~\AA~ in the gas phase equilibrium structure to 0.78~\AA~ in the adsorbed flat BF configurations. This form of binding has been discussed previously\cite{Ataca2009,Singh2012} and is known as a Kubas interaction. More specifically, it arises from stabilization of the 3$d$ state of Ca over the $4s$ state and back-donation of electron density from the valence Ca $d$ state to the $1\sigma^{*}$ state of the \hh molecule. This mechanism is corroborated in the projected density of states (PDOS) of 4H$_2$+Ca@Gr, shown in Fig~\ref{fig:mechdos}(c). \hh bond-weakening can also be found on Sr@Gr and Ba@Gr, indicating that the Kubas mechanism underpins the adsorption of \hh in these systems also. 

Adsorbed \hh molecules on Li, Na, K, Rb, and Cs metal decorated graphene, which are in an upright BF configuration, do not exhibit H--H bond weakening and the effect on the charge density from adsorption is also distinctly different (see SM for the charge density difference  for 4\hh on Na@Gr). Indeed, the PDOS of 4\hh on K@Gr in Fig.~\ref{fig:mechdos}(a) shows no K occupied states near the Fermi energy, indicating complete charge transfer of the K valence electron to graphene and no occupation of the $1\sigma^{*}$ states of the \hh molecules.
Given that alkali adatoms lose their single valence electron to graphene, the resulting positively charged adatom facilitates the adsorption of \hh on the surface through a direct static polarization interaction with \hh molecules. We can see from the adsorption energies in Table 1 that the order of \hh adsorption strength coincides with the polarizing strength of the alkali cation for 3H$_2$ adsorbed, such that the Li adatom  binds \hh the most strongly and the Cs adatom binds \hh the least among the alkali metal adatoms we consider. With more than three \hh molecules adsorbed, the trend holds from Na as Li is small and \hh molecules become sterically hindered. 
%Moreover, 
%addressing Angelos' comment, we have some 4H2-M(+) screening with orca that shows very similar binding energies (differences could be due to computational setups) and the same trend from Li+--->Cs+. However, I think getting into those results might be a bit convoluted here as it would require explaining those calculations also?  

When the adsorption of \hh is very weak, as in the case of Be and Mg decorated graphene, \hh is radially orientated to the adatom while lying flat on graphene. The resulting \hh configuration is similar to \hh physisorption on pristine graphene.\cite{al-hamdani2017cnt} Indeed, it was previously reported that the PBE+D3 adsorption energy of \hh on pristine graphene is $-53$ meV,\cite{al-hamdani2017cnt} while we find that the adsorption energy is $-56$ meV on Mg@Gr. The different adsorption mechanism of \hh on Mg and Be adatoms to other alkaline earth metals can be understood in terms of the metal atom electronic structure. First, the valence $2s$ and $3s$ electrons of Be and Mg, respectively, cannot be stabilized to a $d$ state and therefore they cannot bind \hh molecules via Kubas bonding. Second, the ionization energies of Be and Mg are too high for graphene to oxidize the adatoms. The PDOS of 4H$_2$+Mg@Gr in Fig.~\ref{fig:mechdos}(b) demonstrates the intact Mg valence state and can be seen as an occupied $s$ state just under the Fermi energy of graphene. As a result, Be and Mg remain uncharged atoms that \hh molecules weakly interact with.
\begin{figure}
    \centering
    \includegraphics[width=0.5\textwidth]{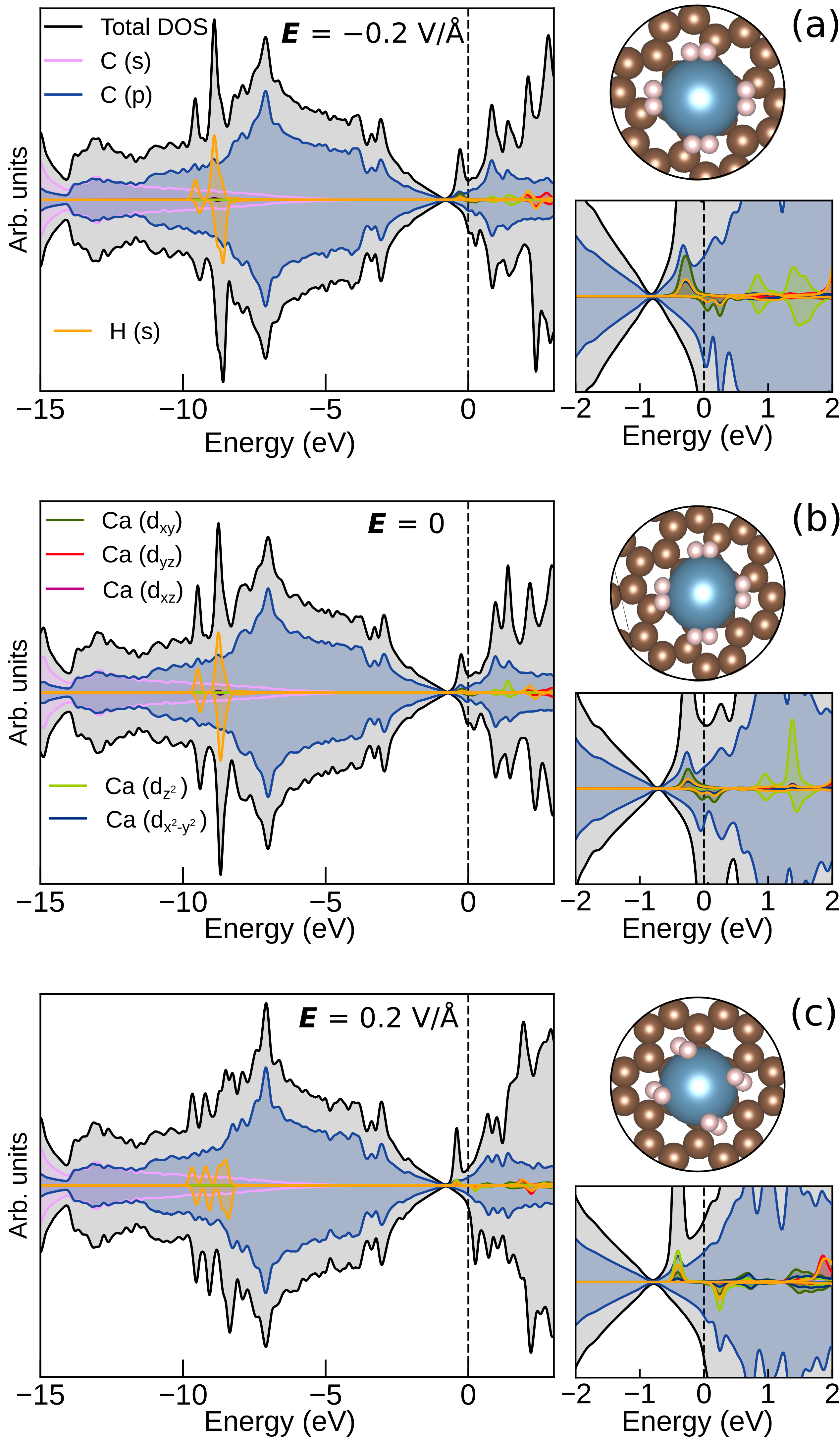}
    \caption{Projected density of states (PDOS) for 4H$_2$+Ca@Gr with external electric fields of $-0.2$ V \AA$^{-1}$ (a,b), no-field (c,d), and +0.2 V \AA$^{-1}$ (e,f). The left panel shows close-ups around the Fermi level (shifted to zero) of the corresponding PDOS on the right. The legend corresponds to all plots. The total DOS (black line, area shaded in grey) is normalized while the projected states are shown only if their contribution is more than $1\%$. The blue shaded regions correspond to C-$p$ states. The projections over spheres centred on the atoms in the unit cell may not add up to the total DOS due to missing interstitial regions. The fully-optimized adsorption structure at each electric field is also shown.}
    \label{fig:efield0}
\end{figure}

\section{Tuning the H$_2$ adsorption energy using an electric field}\label{sec:efield}

\begin{figure}
    \centering
    \includegraphics[width=0.5\textwidth]{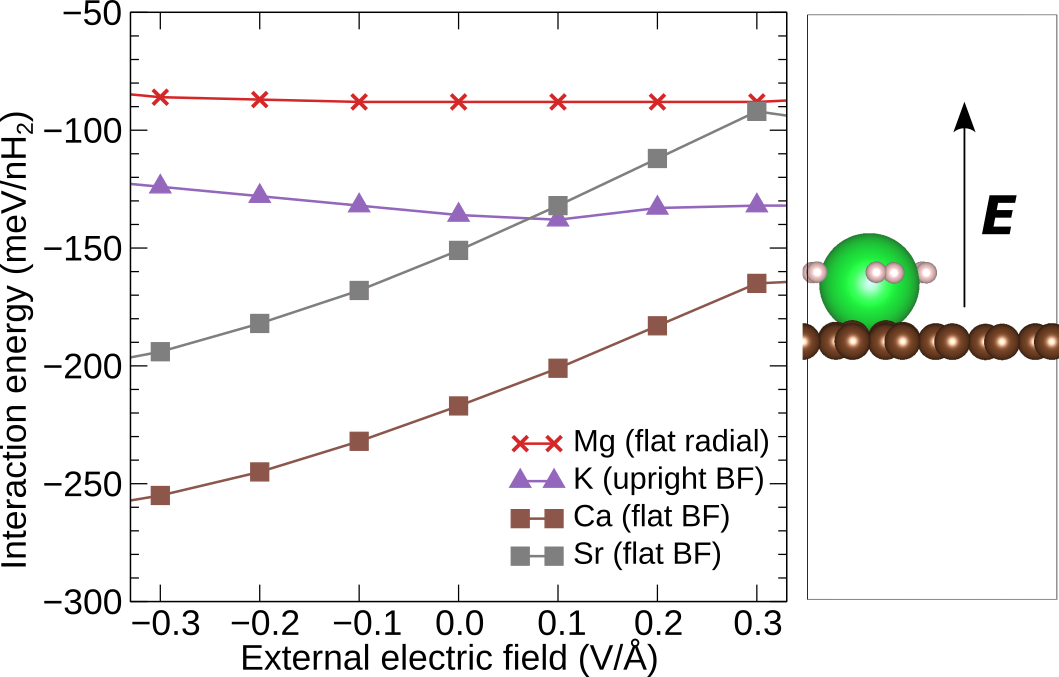}
    \caption{Interaction energy per \textit{unrelaxed} \hh molecule in 4H$_2$+K@Gr (purple triangle), 4H$_2$+Ca@Gr (brown square), 4H$_2$+Sr@Gr (gray square) and 4H$_2$+Mg@Gr (red cross) with respect to the electric force field (in V \AA$^{-1}$). The structures used in calculating the interaction energy are kept fixed at the zero-field adsorbed configuration for each metal element. The interaction energy therefore does not convey the atomically relaxed adsorption energy. The external field was applied in the direction perpendicular to the graphene sheet, as shown in the side panel, and defined in terms of a positive test charge.}
    \label{fig:efield}
\end{figure}

An ideal storage material for \hh would allow the reversible cycling of gas and easy tuning of the \hh adsorption energy would be an additional welcome feature. To this end, we report the effect of applying an electric field on the interaction with \hh bound via the three mechanisms we have established. For the results in Fig.~\ref{fig:efield} we do not allow the atomic positions to relax under the applied electric field and as such, the results indicate the response of only the electron density to an applied field (\textit{i.e.} the high-frequency limit). Specifically, we look at the interaction defined as:
\begin{equation}
    E_{int} = (E^{ads@0}_{ads} - E^{ads@0}_{M@Gr} -  E^{ads@0}_{4H_2} )/ 4
\end{equation}\label{eq:int}
where $E^{ads@0}_{ads}$ is the total energy of the system with 4\hh adsorbed on M@Gr fully optimized at zero-field, while $E^{ads@0}_{M@Gr}$ and $E^{tot}_{4H_2}$ are the total energies of unrelaxed M@Gr and 4\hh in the adsorption configuration at zero-field. Since Eq.~\ref{eq:int} does not take into account any atomic relaxation, the resulting interaction energies do not convey the final adsorption energy at the applied electric field (low frequency limit). For example, it can be seen from Fig.~\ref{fig:efield} that $E_{int}$ at zero-field is lower than $E_{ads}$ reported in Table~\ref{tab:table1} and this is due to the unrelaxed reference subsystems in the definition of $E_{int}$. 
It can be seen from Fig.~\ref{fig:efield} that the effect of an external electric field (applied in the $z$-direction) on the \hh interaction energy with K and Mg decorated graphene is minimal. 
The results suggests that interaction with \hh is not easily perturbed for \hh bound using weak physisorption (flat radial configurations on Be@Gr and Mg@Gr) or static polarization interactions (upright BF on alkali@Gr systems). However, it can be seen from Fig.~\ref{fig:efield} that the \hh molecule interaction with Ca and Sr decorated graphene is strongly affected by an external electric field. With electric fields from $-0.3$ V \AA$^{-1}$ to 0.3 V \AA$^{-1}$ in the $z$-direction, the interaction is decreased by $\sim100$ meV per \hh molecule. Since a positive electric field perpendicular to the graphene sheet draws electrons from the adatom towards graphene, \hh adsorption weakens as the adatom electron density is depleted.  

On relaxation of the 4H$_2$+Ca@Gr system under a positive electric field, we find that the \hh molecules reorient themselves to the upright BF configuration (see Fig.~\ref{fig:efield0}(c)) while the H--H bond length remains elongated (0.78 \AA). This is also reflected in the PDOS of 4H$_2$+Ca@Gr shown in Fig.~\ref{fig:efield0}(c), where the Ca $3d_{xy}$ and $3d_{x^2-y^2}$ states at the valence band edge overlap with \hh $1\sigma^{*}$ state under zero-field and $-0.2$ V \AA$^{-1}$ electric field, whereas under a positive electric field the $3d_{z^2}$ state of Ca is overlapping with \hh $1\sigma^{*}$. In addition, it can be seen that the exchange splitting between the occupied spin-up $3d_{z^2}$ state and the corresponding unoccupied spin-down state is \textit{ca.} 0.5 eV under a positive electric field which indicates single electron occupancy of this state. Under zero or negative electric field, the exchange splitting is smaller ($\sim0.2$ eV) and we see that the corresponding spin-down state is partially occupied. 
In addition, it can be seen from Fig.~\ref{fig:efield0} that occupation of Ca states near the Fermi energy increases with the electric field  decreasing (\textit{i.e.} from +0.2 to $-0.2$ V \AA$^{-1}$). 
This corroborates that there is a higher density of electrons around the Ca adatom under zero and negative electric fields, facilitating a stronger Kubas interaction with \hh molecules. By relaxing a single gas phase hydrogen molecule and the Ca@Gr substrate at $-0.2$ and 0.2 V \AA$^{-1}$ electric force fields (along the same $z$-direction), we find that the adsorption energy of 4\hh on Ca@Gr is $-211$ and $-167$ meV, respectively, per \hh molecule. The difference of $\sim 50$ meV in \hh adsorption energy on Ca@Gr when applying $-0.2$ and 0.2 V \AA$^{-1}$ electric force fields is consistent with the difference in the interaction energy reported in Fig.~\ref{fig:efield}.

\section{Conclusion}\label{sec:conclusions}
We predicted \hh adsorption energies and structures on alkali and alkaline earth metal decorated graphene materials to understand how these substrates can facilitate \hh adsorption. We find three distinct adsorption mechanisms which manifest from the electronic structure of the metal adatom. First, alkali metal adatoms act as positive charges interacting with \hh molecules via attractive electrostatic interactions. Under this mechanism, the \hh molecules are upright on graphene, exposing the most electron-rich bonding region of the \hh molecules to the positively charged adatom. Li@Gr best facilitates this mechanism of binding, with an adsorption energy of $-187$ meV per \hh molecule in 3H$_2$+Li@Gr. Second, small alkaline earth metal atoms, Be and Mg, retain their gas phase electronic structure when adsorbed on graphene and have a negligible impact on adsorbing \hh molecules, leading to weak physisorption. Larger alkaline earth metals, \textit{i.e.} Ca, Sr, and Ba, are partially depleted of valence electron density and more importantly, the $d_{xy}$ and $d_{x^2-y^2}$ states of these atoms are stabilized in favour of the gas phase valence $s$ state. Therefore, in the third mechanism, \hh molecules prefer to bind to the adatoms via Kubas bonding, receiving electron density into the \hh $1\sigma^{*}$ state. This \hh adsorption mechanism is distinguishable due to the resulting elongated H--H bond length. Kubas bonding also results in the strongest adsorption of \hh among the materials we considered, with 4\hh molecules on Ca decorated graphene adsorbing at $-190$ meV per \hh molecule. While this is $\sim10$ meV shy of the adsorption strength we estimate to be necessary for viable \hh storage, a number of factors have not been taken into account at this stage including: zero-point energy vibrations, finite temperature, and importantly, the accuracy of the DFT approximation. In addition, we applied a range of external electric fields to a subset of systems and we find that the adsorption energy of \hh is easily perturbed when \hh molecules are bound via the Kubas interaction. Therefore, it is feasible that a metal decorated graphene system can be made into a viable storage material for hydrogen fuel. More generally, we expect the mechanisms outlined in this work to apply in similar adatom decorated materials. For example, covalent organic frameworks and metal organic frameworks are also promising low-dimensional storage materials, where alkali and alkaline earth metals may play a similar role in binding H$_2$. The experimental synthesis and clear characterization of such materials will be a key step towards the fruition of \hh storage in low dimensional materials. To this end, our results provide some useful indications of which materials to target and what properties can be probed, \textit{e.g.} elongated H--H bonds. In summary, the findings provide a systematic overview of \hh adsorption on alkali and alkaline earth metal decorated graphene and form a basis for developing \hh physisorption storage materials.

\appendix* % asterix specifies that there is only 1 appendix 
\section{Derivation of eq.~\ref{eq:vp-t}.}\label{sec:appendix}
The \hh vapor pressure is a key factor in determining the suitability of \hh storage materials. Theoretical estimations of ideal \hh vapor pressures have been proposed previously,\cite{Li2003,Bhatia2006} resulting in \textit{ca.} $-150$ to $-600$ meV adsorption energy range which is typically considered. The window of adsorption energies ultimately depends on several factors including the choice of fuel cells, device functionality, and the properties of the storage material. In our estimate we considered pressures from 3 bar to 100 bar and temperatures from 270 K to 390 K which covers intermediate and high temperature fuel cells.\cite{Ogungbemi2021,Salam2020} In the following heuristic approach, we show how we evaluate the \hh vapor pressure, using coronene as a model substrate for flat carbon based materials such as graphene, to arrive at our ideal adsorption energy estimate. 
We begin with the Gibbs free energy:
$%\begin{equation}
    G(p,T) = U + pV - TS
$, %\end{equation}
where $p$ is pressure, $T$ is temperature, $U$ is the internal energy, $V$ is volume and $S$ is entropy. The chemical potential, $\mu$, is the Gibbs free energy normalized for the number of particles $N$: 
$%\begin{equation}
\mu(p,T) = {G(p,T) \over N } 
$. %\end{equation}
For the system at equilibrium:
$%\begin{equation}
    \mu_{H_2@Gr} = \mu_{H_2} + \mu_{Gr}
$, %\end{equation}
and we can separate the electronic contribution to the energy, $E_{el}$, which we compute from DFT, leaving the chemical potential of the phase-state (ps) , $\mu^{ps}$:
$%$\begin{equation}
    \mu=E^{el} +\mu^{ps}
$, %\end{equation}
where $E^{el}$ accounts for the electronic energy at 0 K without zero-point energy contributions.
According to Eq.~\ref{eq:ads}, $E_{ads}$ follows from the electronic contributions and thus we can write:
\begin{equation}\label{eq:chempots}
   0 = E_{ads} + \mu^{solid}_{H_2@Gr} - \mu^{gas}_{H_2} - \mu^{solid}_{Gr} 
\end{equation}
where the phase-state is gas for \hh and we assume H$_2$@Gr and Gr are solids. 
As \hh is a homonuclear diatomic gas we assume it here to be ideal such that $\mu^{gas}_{H_2}$ can be expressed as:
\begin{equation}\label{eq:h2mu}
    \mu^{gas}_{H_2} = -k_BT \ln{k_BT\over p \Lambda^3} -k_BT (\ln Z_r + \ln Z_v)
\end{equation}
where $\Lambda = \sqrt{h^2 \over 2 \pi m_{H_2} k_B T}$ is the de Broglie thermal wavelength, $Z_r$ is the rotational partition function and $Z_\nu$ is the vibrational partition function.
As a first approximation 
%\begin{equation}
$
    Z_r \sim {I k_B T \over \hbar^2}
    %Z_r \sim {1\over 2} \left({2 k_BT I \over \hbar^2} + {1\over 3} + {\hbar^2\over 30 I k_BT}\right)
$,
%\end{equation}
where $I = {m_{H_2} d_{H_2}^2 \over 4}$ is the moment of inertia. 
Within the harmonic approximation, the vibrational partition function is  
$%\begin{equation}
    Z_v = %\sum_{i=0}^{\infty} e^{-\beta \hbar \omega_{H_2} (i + 1/2)} =
{\exp(-{\hbar \omega_{H_2} \over 2 k_B T}) \over 1 - \exp(- {\hbar \omega_{H_2} \over k_B T} ) }
$, %\end{equation}
where $\omega_{H_2}$ is the harmonic vibrational frequency of H$_2$.

In the case of solids only phonons need to be considered (in the leading approximation, as the volumes are negligible with respect to the gas phase, so the $pV$ term can be neglected) such that,
\begin{equation}
    \mu^{solid}=-k_BT \ln Z_v^{solid}
\end{equation}
for H$_2$@Gr and Gr, where the vibrational partition function can be evaluated within the harmonic approximation as
$Z_v^{solid} = \prod_{j} {\exp(-{\hbar \omega_j \over 2 k_B T}) \over 1 - \exp(- {\hbar \omega_j \over k_B T} ) }$;
here $\omega_j$ is the vibrational frequency of the $j$-th normal mode. 
H$_2$@Gr has 6 more vibrational modes than Gr due to 5 vibrations from H$_2$ interacting with Gr and 1 mode corresponding to the H$_2$ internal vibration. 
As a leading order approximation, we assume that the vibrations of H$_2$ and Gr are the same in H$_2$@Gr, which allows us to simplify $\mu^{solid}_{H_2@Gr} - \mu^{solid}_{Gr}$ in Eq.~\ref{eq:chempots} as follows:
\begin{equation}\label{eq:solmu}
    \mu^{solid}_{H_2@Gr} - \mu^{solid}_{Gr} +k_B T \ln Z_\nu^{H_2} = -k_BT \ln Z_{i\nu} .
\end{equation}
Here, %$Z_{iv}$ 
$%$\begin{equation}
    Z_{i\nu} = \prod_{j=1}^5 {\exp(-{\hbar \omega_j \over 2 k_BT}) \over 1 - \exp(- {\hbar \omega_j \over k_BT} ) }
$ %\end{equation}. 
is the vibrational partition function for the 5 inter-system modes, having the vibrational frequencies $\omega_j$, $j=1,...,5$.

Thus, by using Eq.~\ref{eq:h2mu} and Eq.~\ref{eq:solmu} in Eq.~\ref{eq:chempots}, we arrive at an expression:
\begin{equation}
-k_BT \ln{k_BT\over p \Lambda^3} -k_BT \ln Z_r = E_{ads} -k_BT \ln Z_{i\nu} 
\end{equation}
From this expression we extract the \hh vapor pressure:
\begin{equation}\label{eq:pres}
p = e^{ { E_{ads} \over k_BT } } {k_BT\over \Lambda^3} {Z_r \over Z_{i\nu}} 
\end{equation}

\begin{figure}
    \centering
    \includegraphics[width=0.5\textwidth]{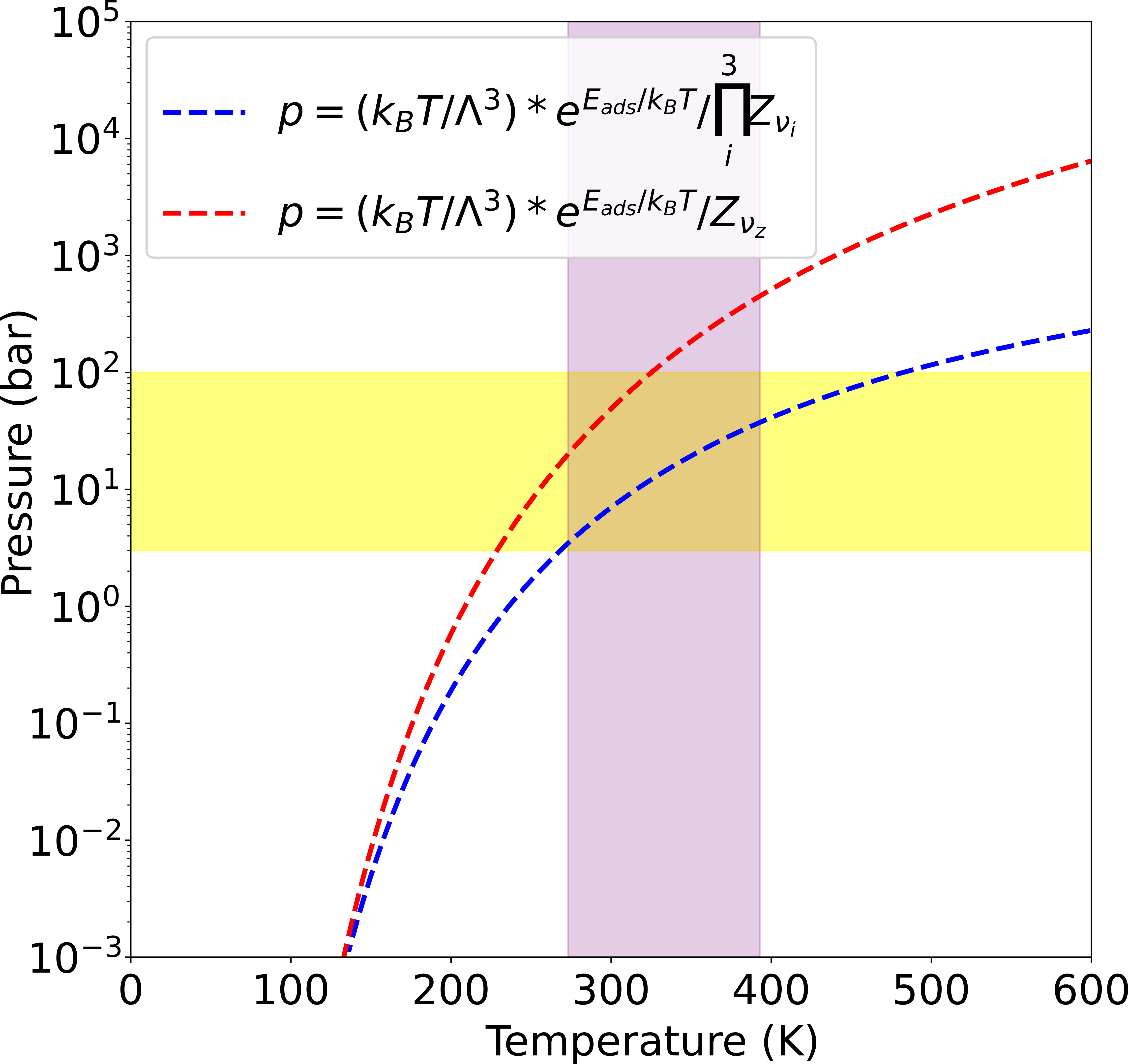}
    \caption{Temperature-pressure adsorption profile for \hh according to the approximations shown in the legend. The red line corresponds to Eq.~\ref{eq:pres_approx4}. An adsorption energy of $-200$ meV is used to demonstrate the effect of different approximations and PBE+D3 inter-system vibrational frequencies of the H$_2$-coronene molecular system.}
    \label{fig:appendix}
\end{figure}

In computing the vapor pressure, we can make a further approximation by assuming that physisorbed \hh rotates freely such that $Z_r$ drops out along with two inter-system vibrational frequencies (which are essentially \hh rotating on the substrate). 
Furthermore, we assume that 2 inter-system vibrations parallel to the surface ($xy$-plane) that are $\sim80$ cm$^{-1}$ are too weak for the harmonic approximation to be useful and thus we can neglect them, 
leaving us with the working equation:
\begin{equation}\label{eq:pres_approx4}
    p \sim {k_BT\over \Lambda^3} e^{ { E_{ads} \over k_BT } } {1\over Z_{v_z}}
\end{equation}
Expanding $\Lambda$, and $Z_{v_z}$ in Eq.~\ref{eq:pres_approx4} yields Eq.~\ref{eq:vp-t}. We consider the effect of this last approximation in Fig.~\ref{fig:appendix} by comparison with using three inter-system vibrations (\textit{i.e.} including those along the $xy$-plane that we deem too weak for the harmonic approximation). We can see that the inclusion of the weak vibrational modes would suggest that even lower adsorption energies could be sufficient at the operating conditions of a fuel cell. 

Finally, it is important to note that we used a molecular system, H$_2$ on coronene, as a model for H$_2$ on pristine graphene, to have an estimate frequency $\omega_z$, which is \textit{ca.} 200~cm$^{-1}$. The ORCA quantum chemistry package,\cite{Neese2020} and the PBE+D3 functional was used to compute vibrational frequencies. For a more accurate pressure-temperature profile, the inter-system vibrational frequencies would need to be known for each substrate material that is considered. Nonetheless, it is interesting that our estimated window of ideal adsorption energy is consistent with previous estimations.\cite{Li2003,Bhatia2006}

\begin{acknowledgments}
Y.S.A. is supported by Leverhulme grant no. RPG-2020-038. A.Z. also acknowledges support by RPG-2020-038. The authors acknowledge the use of the UCL Kathleen High Performance Computing Facility (Kathleen@UCL), and associated support services, in the completion of this work. 
This research used resources of the Oak Ridge Leadership Computing Facility at the Oak Ridge National Laboratory, which is supported by the Office of Science of the U.S. Department of Energy under Contract No. DE-AC05-00OR22725).
A.Z. acknowledge allocation of CPU hours by CSCS under Project ID s1112.
Calculations were also performed using the Cambridge Service for Data Driven Discovery (CSD3) operated by the University of Cambridge Research Computing Service (www.csd3.cam.ac.uk), provided by Dell EMC and Intel using Tier-2 funding from the Engineering and Physical Sciences Research Council (capital grant EP/T022159/1 and EP/P020259/1), and DiRAC funding from the Science and Technology Facilities Council (www.dirac.ac.uk). 
This work also used the ARCHER UK National Supercomputing Service (https://www.archer2.ac.uk), the United Kingdom Car Parrinello (UKCP) consortium (EP/ F036884/1).
\end{acknowledgments}

\bibliography{bibliography.bib}% Produces the bibliography via BibTeX.

\end{document}

\title{Supplemental Material for:\\
Mechanisms of adsorbing hydrogen gas on metal decorated graphene}

\author{Yasmine S. Al-Hamdani}
\affiliation{Department of Earth Sciences, University College London, London WC1E 6BT, United Kingdom}
\affiliation{Thomas Young Centre, University College London, London WC1E 6BT, United Kingdom}
\affiliation{London Centre for Nanotechnology, University College London, London WC1E 6BT, United Kingdom}
\author{Andrea Zen}%
\affiliation{Dipartimento di Fisica Ettore Pancini, Università di Napoli Federico II, Monte S. Angelo, I-80126 Napoli, Italy}
\author{Angelos Michaelides}%
\affiliation{Yusuf Hamied Department of Chemistry, University of Cambridge, Cambridge CB2 1EW, United Kingdom}
\author{Dario Alf\`e}
\affiliation{Dipartimento di Fisica Ettore Pancini, Università di Napoli Federico II, Monte S. Angelo, I-80126 Napoli, Italy}
\affiliation{Department of Earth Sciences, University College London, London WC1E 6BT, United Kingdom}
\affiliation{Thomas Young Centre, University College London, London WC1E 6BT, United Kingdom}
\affiliation{London Centre for Nanotechnology, University College London, London WC1E 6BT, United Kingdom}

\maketitle

\date{\today}

%\tableofcontents 

\section*{Summary}
The supplemental material herein is intended to support ``Mechanisms of adsorbing hydrogen gas on metal decorated graphene'' and provide additional details that may interest some readers. We provide details on: (i) separation distances for loose and tight structural relaxations, (ii) charge density difference plot for the 4H$_2$+Na@Gr system, (iii) H$_2$ dissociation on Ca@Gr, and (iv) example CP2K and VASP input files.

\newpage
\section{Geometry optimizations with ``loose'' and ``tight'' technical parameters}\label{smsec:technical}
In our manuscript we present results from loose DFT geometry relaxations, performed with CP2K for speed and efficiency and the resulting structures were subsequently optimized with tight parameters in VASP (\textit{i.e.} with $9\times9\times1$ \textbf{k}-points and with plane-wave basis sets, avoiding basis set superposition errors). Here, we report the metal adatom to graphene distances as a result of the screening and tight optimizations in Supplementary Table~\ref{stab:mgr}. In Supplementary Table~\ref{stab:h2mgr} we report the average H$_2$ to metal adatom separation distance for 4H$_2$+M@Gr systems. 
\begin{table}[ht]
\caption{\label{stab:mgr} The optimized adatom to graphene distance is reported from CP2K screening with loose technical parameters and subsequent optimization from VASP with tight parameters. The height of the adatom is computed with respect to the average height of carbon atoms in graphene.}
\begin{ruledtabular}
\begin{tabular}{lccccc}
System & {Screening $d_{M-Gr}$ (\AA)} & {Tight $d_{M-Gr}$ (\AA)} & {Absolute change (\AA)} & {$\%$ change} & {$E_{ads}$} \\
Gr+Li  & 1.759                    & 1.704                    & -0.055                              & -3.1                        & -1.279                      \\
Gr+Be  & 3.061                    & 3.218                    & 0.157                               & 5.1                         & -0.181                      \\
Gr+Na  & 2.301                    & 2.189                    & -0.112                              & -4.9                        & -0.719                      \\
Gr+Mg  & 3.216                    & 3.322                    & 0.105                               & 3.3                         & -0.281                      \\
Gr+Ca  & 2.344                    & 2.314                    & -0.030                              & -1.3                        & -0.741                      \\
Gr+K   & 2.599                    & 2.571                    & -0.028                              & -1.1                        & -1.200                      \\
Gr+Rb  & 2.770                    & 2.730                    & -0.039                              & -1.4                        & -1.262                      \\
Gr+Sr  & 2.554                    & 2.497                    & -0.056                              & -2.2                        & -0.753                      \\
Gr+Cs  & 2.909                    & 2.903                    & -0.006                              & -0.2                        & -1.466                      \\
Gr+Ba  & 2.475                    & 2.577                    & 0.103                               & 4.2                         & -1.198                     
\end{tabular}
\end{ruledtabular}
\end{table}
\begin{table}
\caption{\label{stab:h2mgr}The optimized average H$_2$ distance to metal adatom (on graphene) is reported from CP2K screening with loose technical parameters and subsequent optimization from VASP with tight parameters. The distance is computed using the average of 4 H$_2$ molecular distances to the metal where the distances are computed along the vector from the adatom to the middle of the H--H bond for bond-facing molecules and to the nearest H atom in radial configurations.}
\begin{ruledtabular}
\begin{tabular}{lccccc}
System    & {Screening $d_{H_2-M}$ (\AA)} & {Tight $d_{H_2-M}$ (\AA)} & {Absolute change (\AA)} & {$\%$ change}  \\
Gr+Li+4H$_2$ & 1.854                    & 1.789                    & -0.065                              & -3.5                  \\
Gr+Be+4H$_2$ & 2.900                    & 3.182                    & 0.282                               & 9.7                   \\
Gr+Na+4H$_2$ & 2.441                    & 2.258                    & -0.183                              & -7.5                   \\
Gr+Mg+4H$_2$ & 3.292                    & 3.423                    & 0.131                               & 4.0                  \\
Gr+K+4H$_2$ & 2.645                    & 2.571                    & -0.074                              & -2.8                   \\
Gr+Ca+4H$_2$ & 2.299                    & 2.269                    & -0.029                              & -1.3                   \\
Gr+Rb+4H$_2$ & 2.794                    & 2.798                    & 0.004                               & 0.1                  \\
Gr+Sr+4H$_2$ & 2.453                    & 2.466                    & 0.013                               & 0.5                  \\
Gr+Cs+4H$_2$ & 2.935                    & 2.964                    & 0.029                               & 1.0                 \\
Gr+Ba+4H$_2$ & 2.612                    & 2.669                    & 0.057                               & 2.2                 
\end{tabular}
\end{ruledtabular}
\end{table}
%\section{Flat bond-facing and upright bond-facing structure stability}
 
\newpage
\section{Density of states for Na@Gr}\label{smsec:nagr}
\begin{figure}[ht]
    \centering
    \includegraphics[width=1\textwidth]{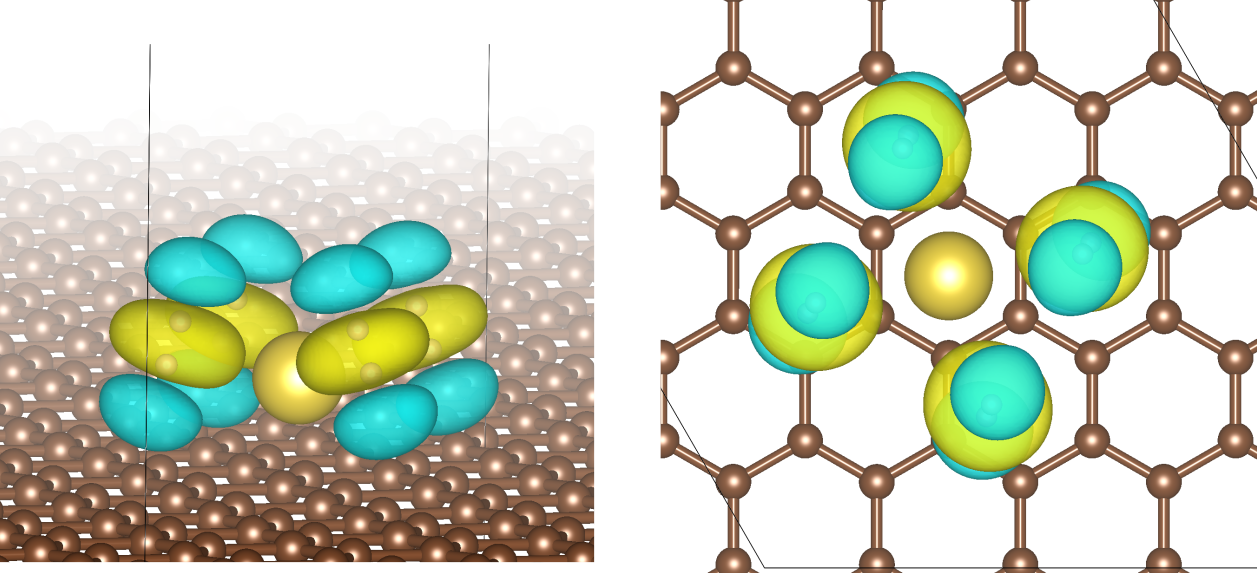}
    \caption{The 4H$_2$+Na@Gr system showing the geometry of H$_2$ molecules around Na and the charge redistribution upon adsorbing H$_2$ molecules.  Charge density difference is shown between 4H$_2$ and Na@Gr using an isosurface level of 0.002 e \AA$^{-3}$. Charge density depletion is shown in blue and charge density accumulation is shown in yellow.}
    \label{sfig:4h2nagr}
\end{figure}

We show the charge density difference for 4H$_2$ adsorbed on Na@Gr in Fig.~\ref{sfig:4h2nagr}. Side-view of the system on the left and top-view on the right. Transparent yellow indicates charge accumulation and blue indicates charge depletion. Carbon atoms in brown,  hydrogen in white, and sodium in yellow. We set the isosurface level at 0.002 e \AA$^{-3}$ for this figure. It can be seen that there is charge accumulation along the H--H bond facing the Na adatom. Moreover, there is no redistribution of charge density within the graphene sheet as a result of H$_2$ adsorption. 

\newpage
\section{Gauging the stability of 4H$_2$+Ca@Gr}\label{smsec:stab}
%Two aspects, among others, 
Among different aspects which can affect the efficiency of a potential storage material with adatoms, 
%is the diffusion of the adatom across the substrate and 
is the dissociation of H$_2$ on the material. This process can be a form of H$_2$ storage known as the spillover effect but it is not the mechanism which we are targeting as it typically involves higher energy costs for desorbing hydrogen. 

%First for the diffusion of Ca across pristine graphene, we used NEB calculations as implemented in the VTST Tools for VASP. Our starting structure is the fully optimized Ca@Gr system where Ca is adsorbed on the hollow of a carbon ring. The final structure is the fully optimized Ca@Gr system with Ca at a neighbouring hollow site. Five images were used with the climbing image algorithm and a spring force constant of 5 eV \AA$^{-2}$ with nudging. Atomic forces were converged to 0.01 eV \AA$^{-1}$ and a $5\times5\times1$ \textbf{k}-point mesh was used. The resulting PBE+D3 barrier for Ca to diffuse is 0.14 eV. 

\begin{figure}[ht]
    \centering
    \includegraphics[width=1\textwidth]{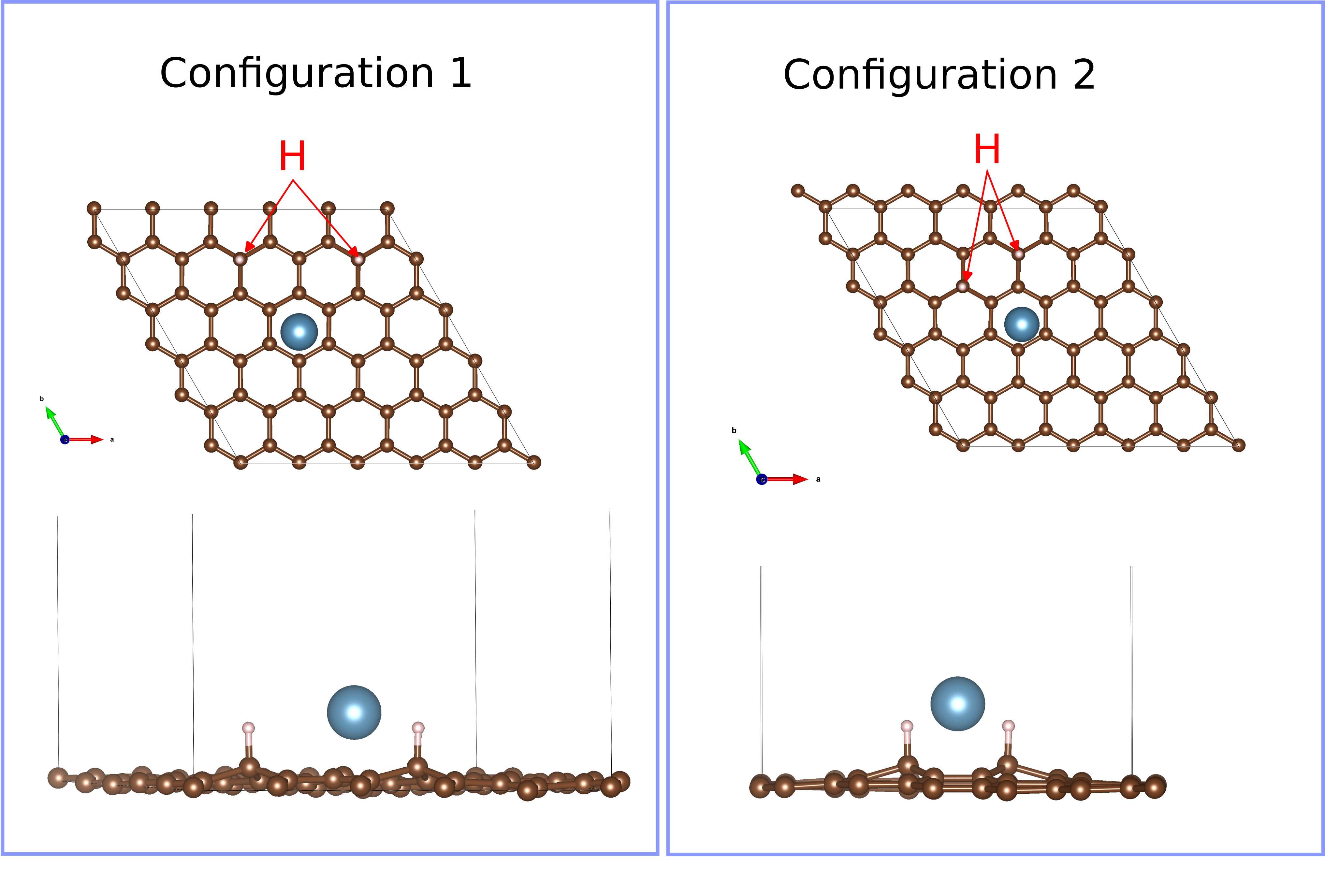}
    \caption{Two fully optimized configurations of 2H+Ca@Gr showing top and side views. Configuration 1 was found to be 2.19 eV less stable than H$_2$+Ca@Gr and configuration 2 was found to be 1.75 eV less stable than H$_2$+Ca@Gr.}
    \label{smfig:diss}
\end{figure}

%Second, 
To considered the dissociative adsorption of H$_2$ on Ca@Gr, we attempted to adsorb 2H on the Ca adatom, but despite several starting configurations, none converged successfully. However, we also considered H atoms adsorbing on graphene directly (in the vicinity of Ca) and found that these structures were stable. See Fig.~\ref{smfig:diss} for the final optimized configurations of dissociatively adsorbed hydrogen on Ca@Gr. The total energies of these structures indicates that they are more than 1.7 eV less stable than the physisorbed H$_2$+Ca@Gr system (which was fully optimized using the same settings, \textit{i.e.}  atomic forces were converged to 0.01 eV \AA$^{-1}$ and a $5\times5\times1$ \textbf{k}-point mesh was used.

\newpage
\section{Input files for CP2K and VASP}\label{smsec:input}
Some example input files are given here to help guide the interested reader. The CP2K input for the screening optimization of 4$H_2$+Ca@Gr is as follows:
\begin{verbatim}
 &GLOBAL
   RUN_TYPE  GEO_OPT
 &END GLOBAL
 &MOTION
   &GEO_OPT
     OPTIMIZER  BFGS
     MAX_ITER  1000
     MAX_DR     3.0000000000000001E-03
     MAX_FORCE     1.0000000000000000E-04
     &BFGS
     &END BFGS
   &END GEO_OPT
   &PRINT
     &TRAJECTORY  SILENT
       FORMAT  PDB
     &END TRAJECTORY
   &END PRINT
 &END MOTION
 &FORCE_EVAL
   METHOD  QS
   &DFT
     BASIS_SET_FILE_NAME BASIS_MOLOPT
     POTENTIAL_FILE_NAME POTENTIAL
     UKS  T
     CHARGE  0
     &SCF
       MAX_SCF  60
       EPS_SCF     1E-06
       SCF_GUESS  ATOMIC
       &OT  T
         MINIMIZER  DIIS
         PRECONDITIONER  FULL_KINETIC
       &END OT
       &OUTER_SCF  T
         EPS_SCF     1E-06
       &END OUTER_SCF
       &PRINT
         &RESTART  OFF
         &END RESTART
       &END PRINT
     &END SCF
     &QS
       EPS_DEFAULT     1.0000000000000000E-10
       EXTRAPOLATION  ASPC
       EXTRAPOLATION_ORDER  3
       METHOD  GPW
     &END QS
     &MGRID
       NGRIDS  5
       CUTOFF     300
       REL_CUTOFF     30
     &END MGRID
     &XC
       DENSITY_CUTOFF     1.0000000000000000E-10
       GRADIENT_CUTOFF     1.0000000000000000E-10
       TAU_CUTOFF     1.0000000000000000E-10
       &XC_FUNCTIONAL  NO_SHORTCUT
         &PBE  T
         &END PBE
       &END XC_FUNCTIONAL
       &VDW_POTENTIAL
         POTENTIAL_TYPE  PAIR_POTENTIAL
         &PAIR_POTENTIAL
           R_CUTOFF     1.6000000000000000E+01
           TYPE  DFTD3
           PARAMETER_FILE_NAME dftd3.dat
           REFERENCE_FUNCTIONAL PBE
           CALCULATE_C9_TERM  T
         &END PAIR_POTENTIAL
       &END VDW_POTENTIAL
     &END XC
     &POISSON
       PERIODIC  XYZ
     &END POISSON
     &PRINT
     &END PRINT
   &END DFT
   &SUBSYS
     &CELL
       A     1.2324999999999999E+01    0.0000000000000000E+00    0.0000000000000000E+00
       B    -6.1624999999999970E+00    1.0673763101643207E+01    0.0000000000000000E+00
       C     0.0000000000000000E+00    0.0000000000000000E+00    2.0000000000000004E+01
       MULTIPLE_UNIT_CELL  1 1 1
     &END CELL
     &COORD
C   -1.5070046352432811E-03   -2.8594858403717814E-03    8.7264662134649687E-02
C   -6.1588502213592317E-03    1.4174383600048301E+00    1.1530598278636271E-01
C    2.4634661050903794E+00   -8.5453292661629744E-03    1.1685302644076113E-01
C    2.4557303549054690E+00    1.4149111057822412E+00    1.7092057120160772E-01
C    1.1090009319624890E+01    2.1330613811293331E+00    1.0862553674620437E-01
C   -1.2387711544127211E+00    3.5658405563845554E+00    1.1520798322934885E-01
C    1.2303313818183792E+00    2.1227143973543585E+00    1.6888555500467917E-01
C    1.2212641565360645E+00    3.5649872472451505E+00    2.4676087474853667E-01
C    4.9405579607871157E+00   -1.7115700517701452E-03    1.1564922145965990E-01
C    4.9423347486023106E+00    1.4170451686908940E+00    1.6705364572196682E-01
C    9.8577607962974891E+00    4.2724403112180855E+00    9.4832881934529656E-02
C   -2.4724773246572740E+00    5.6975497163013351E+00    9.7086238900332147E-02
C    2.4756674674064589E+00    4.2789286708626904E+00    1.7063376871666172E-01
C    2.4678095280870700E+00    5.7023812824738638E+00    1.1710532608002147E-01
C    7.4036928438374208E+00   -3.8619428387756329E-03    9.7466979817644128E-02
C    7.3984729667613758E+00    1.4213556961443254E+00    9.5099031628462630E-02
C    8.6261442238846673E+00    6.4057943739004797E+00    9.2056806517922987E-02
C   -3.7028495170350806E+00    7.8276060300281145E+00    9.0232771914782975E-02
C    3.6999721469952189E+00    6.4084999859668335E+00    8.8286353863463687E-02
C    3.6986910302226224E+00    7.8313139383956756E+00    8.5569331793162701E-02
C   -1.0994645819299672E-02    4.2767113027831893E+00    1.6678002590348903E-01
C   -9.3221054028297355E-03    5.6955278445815702E+00    1.1562882112165268E-01
C    3.7101407004189184E+00    2.1288995503226622E+00    2.4673624014559625E-01
C    3.7009796806889521E+00    3.5711208846320077E+00    1.6882503890882192E-01
C   -1.2351482210072493E+00    6.4058609902544150E+00    9.4476801794855389E-02
C   -1.2318393361588096E+00    7.8286289653454872E+00    8.7948667379910681E-02
C    6.1701206317802386E+00    2.1278990652321976E+00    1.1566949699831119E-01
C    6.1662654534291219E+00    3.5607180757439658E+00    1.0944182092345275E-01
C    1.2337400754045347E+00    6.4088294336475604E+00    1.0972876189108031E-01
C    1.2348717506072437E+00    7.8254450422562911E+00    9.2997718923721343E-02
C    4.9375208713480809E+00    4.2763072858835249E+00    1.1582365903753196E-01
C    4.9328350330611794E+00    5.6965950899415017E+00    8.8171400318280524E-02
C    9.8633134240164750E+00   -4.5770663229927613E-03    9.0376407277156109E-02
C    9.8636233310804844E+00    1.4234975976393980E+00    9.2621553973285031E-02
C    8.6287941077688863E+00    2.1356602297873724E+00    8.8892472129754055E-02
C    8.6274492552379325E+00    3.5580276120771406E+00    8.8950576537178028E-02
C    7.3926656187175421E+00    4.2702415379737566E+00    9.3184257431316864E-02
C    7.3929464441445525E+00    5.6982763009971187E+00    9.0773661566221756E-02
C    6.1656091890990705E+00    6.4068730696409641E+00    8.5559561875345136E-02
C    6.1598300210510875E+00    7.8298203776014335E+00    8.1224805132829286E-02
C   -4.9299217386073160E+00    8.5361451041611112E+00    8.4934330211961048E-02
C   -4.9311899706977291E+00    9.9590199919807976E+00    8.7469819985139710E-02
C   -2.4661015232872283E+00    8.5420625369473147E+00    9.2155452095529014E-02
C   -2.4649680765768700E+00    9.9586424992595006E+00    1.0911030686350123E-01
C    6.0877549634965765E-04    8.5388581180536356E+00    8.8324090839118213E-02
C    3.9726144072097288E-03    9.9616509904493427E+00    9.4754166597754541E-02
C    2.4715422165071219E+00    8.5399097168572027E+00    9.0904097904078593E-02
C    2.4675686659684937E+00    9.9616939268659763E+00    9.2375389456563345E-02
C    4.9338549824888913E+00    8.5377017696053947E+00    8.1265271560140240E-02
C    4.9280954047061645E+00    9.9606780518049973E+00    8.4936925850554498E-02
Ca    2.4664538136327154E+00    2.8477669026941124E+00    2.4102328488779223E+00
H    2.4836835950164385E-01    2.6541792671418420E+00    2.9896081546994528E+00
H    4.5021038141290654E-01    1.9078618873019026E+00    3.0226699857022572E+00
H    1.5053727137023263E+00    4.8573703669349095E+00    2.9837015525210706E+00
H    2.2508998950193999E+00    5.0617714589938085E+00    3.0185808716812232E+00
H    3.4341951106382758E+00    8.4192389778639176E-01    2.9902525437595062E+00
H    2.6897332741735047E+00    6.3414179957178385E-01    3.0252997541521176E+00
H    4.6838569890055890E+00    3.0540596198883803E+00    2.9907110362917808E+00
H    4.4771158176259593E+00    3.7993740824135909E+00    3.0172052175729469E+00
     &END COORD
     &KIND H
       BASIS_SET DZVP-MOLOPT-SR-GTH
       POTENTIAL GTH-PBE-q1
       &POTENTIAL
 1
  0.2000000000000000E+00 2 -0.4178900440000000E+01  0.7244633100000000E+00
 0
         # Potential name:  GTH-PBE-Q1  for symbol:  H
         # Potential read from the potential filename: POTENTIAL
       &END POTENTIAL
     &END KIND
     &KIND B
       BASIS_SET DZVP-MOLOPT-SR-GTH
       POTENTIAL GTH-PBE-q3
     &END KIND
     &KIND C
       BASIS_SET DZVP-MOLOPT-SR-GTH
       POTENTIAL GTH-PBE-q4
       &POTENTIAL
 2 2
  0.3384712400000000E+00 2 -0.8803673979999999E+01  0.1339210850000000E+01
 2
  0.3025757500000000E+00 1  0.9622486650000001E+01
  0.2915069400000000E+00 0
         # Potential name:  GTH-PBE-Q4  for symbol:  C
         # Potential read from the potential filename: POTENTIAL
       &END POTENTIAL
     &END KIND
     &KIND N
       BASIS_SET DZVP-MOLOPT-SR-GTH
       POTENTIAL GTH-PBE-q5
     &END KIND
     &KIND O
       BASIS_SET DZVP-MOLOPT-SR-GTH
       POTENTIAL GTH-PBE-q6
     &END KIND
     &KIND F
       BASIS_SET DZVP-MOLOPT-SR-GTH
       POTENTIAL GTH-PBE-q7
     &END KIND
     &KIND Na
       BASIS_SET DZVP-MOLOPT-SR-GTH
       POTENTIAL GTH-PBE-q9
     &END KIND
     &KIND Li
       BASIS_SET DZVP-MOLOPT-SR-GTH
       POTENTIAL GTH-PBE-q3
     &END KIND
     &KIND K
       BASIS_SET DZVP-MOLOPT-SR-GTH
       POTENTIAL GTH-PBE-q9
     &END KIND
     &KIND Mg
       BASIS_SET DZVP-MOLOPT-SR-GTH
       POTENTIAL GTH-PBE-q10
     &END KIND
     &KIND Ca
       BASIS_SET DZVP-MOLOPT-SR-GTH
       POTENTIAL GTH-PBE-q10
       &POTENTIAL
 4 6
  0.3900000000000000E+00 2 -0.4167072220000000E+01 -0.1583797810000000E+01
 3
  0.2893557200000000E+00 2  0.2053187635000000E+02 -0.7129785780000000E+01
  0.9204513870000000E+01
  0.3278820600000000E+00 2  0.5805605130000000E+01 -0.4287533600000000E+00
  0.5073078200000000E+00
  0.6796171300000000E+00 1  0.5806826000000000E-01
         # Potential name:  GTH-PBE-Q10  for symbol:  CA
         # Potential read from the potential filename: POTENTIAL
       &END POTENTIAL
     &END KIND
     &TOPOLOGY
       COORD_FILE_NAME geo.xyz
       COORD_FILE_FORMAT  XYZ
       NUMBER_OF_ATOMS  59
       MULTIPLE_UNIT_CELL  1 1 1
     &END TOPOLOGY
   &END SUBSYS
 &END FORCE_EVAL
\end{verbatim}

An example VASP input file for the optimization of structures:
\begin{verbatim}
ISTART = 0
ENCUT = 500
EDIFF = 1E-6
ALGO = Fast
PREC = Accurate
NELM = 125   
ADDGRID = .TRUE.    
LASPH   = .TRUE.    
LREAL   = FALSE
ISMEAR = 0     
SIGMA = 0.05  
ISPIN = 2  
IBRION = 2   
NSW = 1000   
ISIF = 0     
EDIFFG = -1E-2
NWRITE = 2
LWAVE = .FALSE.
NPAR = 4
GGA    = PE   
IVDW = 11
\end{verbatim}